\documentclass[11pt]{article}

\newcommand{\be}{\begin{equation}}
\newcommand{\ee}{\end{equation}}

\title{\bf Counting the microstates of a vacuum black ring} \author{{\bf 
Harvey S. Reall} \\ Department of Applied Mathematics and Theoretical 
Physics \\ Centre for Mathematical Sciences \\ Wilberforce Road, Cambridge 
CB3 0WA, UK \\ hsr1000@cam.ac.uk}

\date{19 December 2007}                                           

\begin{document}
\maketitle

\begin{abstract}

The Bekenstein-Hawking entropy of an extremal vacuum black ring is derived 
from a microscopic counting of states. The entropy of extremal 
Kaluza-Klein black holes with ergospheres is also derived.

\end{abstract}

\section{Introduction}

There has been recent progress in using string theory to provide a 
microscopic calculation of the entropy of extremal
vacuum black holes \cite{EH,EM,HR}. In particular, Horowitz and Roberts 
have shown how the Bekenstein-Hawking entropy of an extremal Kerr black 
hole \be \label{eqn:extremalkerr}
 S = 2 \pi |J|
\ee
can been reproduced from a statistical counting of microstates \cite{HR}. 
Extremality is important in these calculations since extremal black holes obey an attractor mechanism (see \cite{senreview} for a review), even when rotating \cite{rotatt}, which explains why the entropy of such black holes does not change as the string coupling is decreased. This implies that the entropy calculated from a solution of classical gravity can be compared directly with the entropy calculated microscopically \cite{agmdab}.

In this paper, I shall extend these calculations to a different class of 
vacuum black 
holes: black rings. Black rings with a single non-vanishing angular 
momentum were constructed in \cite{ER} but they do not admit a regular 
extremal limit. However, black rings with two angular momenta, constructed 
in \cite{PS}, do.\footnote{Some physical properties of the solutions of 
\cite{PS} have been discussed in \cite{henriette}.} An extremal 
vacuum 
black ring has two parameters: the two angular momenta $J_1$, $J_2$. 
The present paper is motivated by the observation that
the  Bekenstein-Hawking entropy of an extremal vacuum black ring is \be 
\label{eqn:extremalring}
 S = 2\pi |J_2|. \ee The similarity with equation (\ref{eqn:extremalkerr}) 
suggests that a microscopic derivation of this result may be possible.

The similarity of equations (\ref{eqn:extremalkerr}) and 
(\ref{eqn:extremalring}) arises from the fact that an extremal black ring 
has a near-horizon geometry that is isometric to the near-horizon geometry 
of an extremal boosted Kerr string with $J=J_2$ \cite{KLR}. The entropy of 
the latter is independent of the boost, and hence equals the entropy of an 
unboosted Kerr string. Upon dimensional reduction, this is just the 
entropy of an extremal Kerr black hole.

The idea that we shall exploit in this paper arises from the study of BPS 
black rings \cite{bpsrings}, for which succesful microscopic calculations 
of the entropy have been performed \cite{benakraus,cyrier}. Since black 
rings can be regarded as rotating loops of black string, these 
calculations start by assuming that the low-energy dynamics of a BPS black 
ring should be described by the CFT that governs the low energy dynamics 
of the corresponding BPS black string. More precisely, the calculations 
involve identifying the charges of a BPS black ring with the charges of a 
BPS boosted black string wrapped on a Kaluza-Klein circle, and then 
calculating the microscopic entropy of the latter. One might expect this 
approach to work for "skinny rings", for which the radius $R_1$ of the 
$S^1$ (of the $S^1 \times S^2$ horizon) is much greater than the radius 
$R_2$ of the $S^2$. Indeed, for extremal dipole rings one obtains the 
correct result for the entropy calculated this way for large $R_1/R_2$ 
\cite{dipole}. For BPS rings, it turns out that this microscopic calculation 
correctly reproduces the Bekenstein-Hawking entropy for {\it arbitrary} 
$R_1/R_2$  \cite{benakraus,cyrier}. (In fact, the entropy of extremal 
dipole rings can also be calculated for arbitrary $R_1/R_2$ 
\cite{emparannew}.)

For extremal vacuum rings, we shall see that $R_1/R_2 \sim J_1/J_2$. Hence 
we might expect the above method to work for large $J_1/J_2$. However, the 
entropy (\ref{eqn:extremalring}) is independent of $J_1$. Hence it is 
independent of $R_1/R_2$. Phrasing things differently, the leading term in 
the expansion of the entropy in large $R_1/R_2$ (i.e. large $J_1/J_2$) is 
exact. This is an encouraging sign that a microscopic state counting based 
on regarding the black ring as a boosted black string, which works so well 
for BPS rings, may also work for extremal vacuum rings with arbitrary 
$J_1/J_2$.

The idea, then, is to take the microscopic theory of the black ring to be 
the theory governing an extremal boosted Kerr black string. This is the 
theory used for the Kerr microstate counting in \cite{HR}. We need to map 
the charges $J_1,J_2$ of the black ring to the black string charges. For 
BPS black rings, there is disagreement over how to do this, with two 
different methods proposed \cite{benakraus,cyrier}. However, in the vacuum 
case studied here, it seems quite clear cut: the isometry between the 
black ring and black string near-horizon geometries fixes the 
identification uniquely. In any case, the only result we need to do the 
calculation is the identification of the black string angular momentum $J$ 
with the black ring angular momentum $J_2$, which looks uncontroversial.

Our microscopic calculation, which is a slight modification of \cite{HR} 
also allows us to extend the results of \cite{EM} governing "ergo-branch" 
Kaluza-Klein black holes to arbitrarily high angular momentum.

This paper is organized as follows. In section 2, we present some 
properties of extremal vacuum black rings. In section 3, we use the 
isometry between near-horizon geometries to determine the charges of the 
boosted black string that we will use in the entropy calculation. Section 
4 contains the entropy calculation. Section 5 contains a brief discussion.

\section{Extremal black ring}

An extremal vacuum black ring is specified by two parameters $k>0$ and $0 < \lambda <2$. $k$ has dimensions of length and sets a scale for the solution. $\lambda$ is dimensionless. The mass is
\be
 M = \frac{12 k^2 \pi \lambda}{G_5 (2-\lambda)^2},
\ee
and the angular momenta are (choosing them to be positive)
\be
 J_1 = \frac{8 k^3 \pi \lambda ( 4 + 8 \lambda + \lambda^2)}{G_5 (2-\lambda)^3 (2 + \lambda)}, \qquad J_2 = \frac{32 k^3 \pi \lambda^2}{G_5 (2-\lambda)^3 (2+\lambda)}.
\ee
The solution is uniquely determined by its conserved charges, in constrast with non-extremal rings. To see this, note that
\be
 \frac{J_2}{J_1} = \frac{4\lambda}{4 + 8\lambda + \lambda^2}.
\ee
The function on the RHS is monotonically increasing for $0< \lambda <2$. Hence $\lambda$ is uniquely determined by $J_2/J_1$, and we have $0<J_2/J_1<1/3$, i.e.,
\be
\label{eqn:Jrange}
 J_1 > 3J_2 >0. \ee Having fixed $\lambda$, $k$ is uniquely specified by 
the value of, say, $J_1$. Hence the solution is uniquely specified by 
$(J_1,J_2)$ in the range (\ref{eqn:Jrange}). In the limit $J_1/J_2 
\rightarrow 3$, the solution probably becomes an extremal Myers-Perry 
\cite{myersperry}
solution.\footnote{Although I have not checked this. Evidence in favour of 
this comes from comparing the mass of an extremal ring to the mass of an 
extremal MP solution with the same angular momenta: $M_{MP}^3 = 27\pi (J_1 
+ J_2)^3/(32G_5)$. One finds that $M_{ring}/M_{MP}$ is a monotonic 
increasing function of $J_2/J_1$, attaining its maximum value of $1$ as 
$J_2/J_1 \rightarrow 1/3$. Hence the masses of the solutions agree in this 
limit. The ratio of the entropies is $S_{ring}/S_{MP} = (J_2/J_1)^{1/2} < 
1/\sqrt{3}$, so entropy would be discontinuous in the limit in which the 
ring became a MP solution, just as happens in the limit in which a BPS 
black ring approaches a topologically spherical black hole.} Eliminating 
$k$ and $\lambda$ from $M$ gives \be
 M^3 = \frac{27 \pi}{4G_5} J_2 ( J_1 - J_2).
\ee
Equation (\ref{eqn:Jrange}) implies that
\be
 \frac{27 \pi}{2 G_5}J_2^2 < M^3 < \frac{3 \pi}{2G_5} J_1^2.
\ee
At the horizon, the radius of the $S^1$ varies over the $S^2$. At the poles of the $S^2$, the $S^1$ has radius
\be
 R_1 = \frac{2k(2+\lambda)}{2-\lambda}.
\ee
(At the equator, the radius of the $S^1$ is $\sqrt{3/2}$ times larger.) This can be rewritten as
\be
 \frac{R_1^3}{G_5} = \frac{4 (J_1 - J_2)^2}{\pi J_2}.
\ee
The $S^2$ is not homogeneous, but we can define an effective radius $R_2$ by saying that it has area $4\pi R_2^2$. This gives 
\be
 R_2 = \frac{4k\lambda}{4-\lambda^2},
\ee
which implies
\be
 \frac{R_2^3}{G_5} = \frac{J_2^2}{2 \pi (J_1-J_2)},
\ee
Note that
\be
 \frac{R_1}{R_2} = \frac{2(J_1-J_2)}{J_2} \ee hence extremal rings with 
$J_1 \gg J_2$ are skinny whereas extremal rings with $J_1 \sim 3J_2$ are 
fatter with $R_1/R_2\sim 4$.

In order to neglect higher-derivative corrections, we need $R_1$ and $R_2$ 
to be large in Planck units, which requires $J_2^2 \gg J_1 - J_2 \gg 
\sqrt{J_2} \gg 1$.

\section{Matching to a Kerr string}

Take the product of the 4d Kerr solution with a flat direction, boost in this direction, compactify this direction into a circle and then take the extremal limit. This gives the 3-parameter extremal boosted Kerr black string solution. The 3 parameters are the angular momentum $J$, the number $N_0$ of units of momentum around the KK circle, and the asymptotic radius $R$ of this circle. 

The near-horizon geometry of an extremal vacuum black ring was obtained in \cite{KLR}. It was shown that this is globally isometric to the near-horizon geometry of an extremal boosted Kerr black string. In order the make the correspondence between black ring and black string precise, we need to relate the 3 parameters of the black string to the 2 parameters of the black ring. The desired relation follows from the isometry between the near-horizon geometries. One finds that
\be
  J=J_2,
\ee  
\be
\label{eqn:N0J}  
  N_0 = J_1 - J_2,
\ee
and $R=R_1/\sqrt{2}$, so
\be
 \frac{R^3}{G_5} = \frac{\sqrt{2} (J_1 - J_2)^2}{\pi J_2}.
\ee
We should note that, for BPS rings, there is disagreement in the 
literature over how the parameters of the black string should be related 
to those of the black ring \cite{benakraus,cyrier}. In particular, there 
is disagreement over the value of $N_0$ for BPS rings.
For vacuum rings, the above argument seems clear cut (and appears 
to favour the 
proposal of \cite{benakraus} over that of \cite{cyrier}) but 
it doesn't generalize to BPS rings since the near-horizon solution of the 
latter contains fewer parameters than the full solution hence matching 
near-horizon solutions does not allow one to match uniquely parameters in 
the full solution. Even if one disagrees with the above value for $N_0$, 
the argument below is independent of the precise value of $N_0$ (because 
the entropy doesn't depend on $N_0$).
 
\section{Entropy calculations}

\subsection{Kaluza-Klein black holes}

The boosted Kerr black string can be dimensionally reduced to give an 
extremal 4d Kaluza-Klein black hole. This solution is specified by its 
electric charge $N_0$, which is just the number of units of momentum 
around the KK circle, and by its angular momentum $J$. Taking the product 
of this solution with a 6-torus and interpreting the KK circle as the 
M-theory circle, this solution carries D0-brane charge $N_0$. More general 
extremal KK black holes \cite{KKBH} are parameterized by 
$(N_0,N_6,J)$ where $N_6$ is 
KK monopole charge in 11 dimensions, or equivalently D6-brane charge in 10 
dimensions. Such black holes fall into two classes. In the terminology of 
\cite{rotatt}, the "ergo-free branch" of black holes has $J^2 < N_0^2 
N_6^2/4$ and the entropy \be \label{eqn:ergofree} S_{\rm ergo-free} = 2\pi 
\sqrt{N_0^2 N_6^2/4 - J^2} \ee
 of such black holes was calculated in \cite{EH} by dualizing to a non-BPS 4-charge configuration and arguing that results derived in the BPS case could be extended to this case. "Ergo-branch" black holes have $J^2 > N_0^2 N_6^2/4$ and the entropy 
\be
\label{eqn:ergo}
 S_{\rm ergo}= 2\pi \sqrt{J^2-N_0^2N_6^2/4}
\ee 
 of such black holes was calculated in \cite{EM} assuming that 
\be
\label{eqn:consistency}
 1 - N_0^2 N_6^2/(4J^2) \ll 1.
\ee 
This condition arises from requiring that the dual 4-charge configuration admit an $AdS_3$ factor in its decoupling limit, so that CFT arguments are legitimate. The Kerr string has $N_6 = 0$ so it is on the ergo-branch but does not satisfy the condition (\ref{eqn:consistency}). This problem was circumvented in \cite{HR} by an ingenious transformation that interchanges the ergo and ergo-free branches of solutions. 

\subsection{The method of Horowitz and Roberts}

The argument of \cite{HR} involved two novel steps that we shall exploit below.

{\it Covering spaces.} Consider an extremal KK black hole with parameters $(N_0,N_6,J)$, KK circle radius $R$ and entropy $S$. Assume $N_6>0$, so the KK circle is non-trivally fibered over the 4d spacetime. The topology of the horizon (or spatial infinity) is $S^3/Z_{N_6}$. If $K$ divides $N_6$ then one can pass to a $K$-fold covering space of this solution, keeping the local geometry fixed in 11d Planck units. The new parameters are $(N_0 K^2, N_6/K, KJ)$ and the KK circle has radius $KR$ \cite{HR}. Working in the covering space amounts to considering $K$ copies of the original black hole, so the entropy becomes $KS$. 

{\it Branch exchange.} Consider a KK black hole with parameters $(N_0,1,J)$, KK circle radius $R$ and entropy $S$. Let $R \rightarrow \infty$. This gives an asymptotically flat\footnote{
If $N_6 >1$ then the solution would not be asymptotically flat.}
extremal Myers-Perry \cite{myersperry} black hole \cite{EM}. The angular momenta in orthogonal planes are $J_{1,2} = N_0/2 \pm J$ \cite{EM}. Now perform a reflection to change the sign of $J_2$. This has the effect of interchanging $N_0$ and $2J$. Finally, extrapolate back to finite $R$. The new solutions has parameters $(2J,1,N_0/2)$. Hence if the original solution was on the ergo branch then the new solution is on the ergo-free branch and vice-versa. The attractor mechanism ensures that the entropy does not change as $R$ is varied, and a reflection clearly does not change the entropy. Hence the final solution must have the same entropy as the initial solution, as can be checked using (\ref{eqn:ergofree}) and (\ref{eqn:ergo}).\footnote{
Note that the point of this argument is that is explains why the solutions with charges $(N_0,1,J)$ and $(2J,1,N_0/2)$ have the same entropy, which would otherwise be a mysterious coincidence.} However, there is no reason for the mass to be invariant and indeed it is not (explicit expressions for the mass are given in \cite{EM}).

\subsection{Black ring}

We have argued above that calculating the entropy of an extremal vacuum 
black ring should be equivalent to calculating the entropy of an extremal 
boosted Kerr string. Hence our starting point is the extremal boosted Kerr 
string, or extremal KK black hole, with parameters $(N_0,0,J)$. Let $S$ 
denote the entropy of this solution. T-dualizing this on the entire $T^6$ 
gives a 
solution with parameters $(0,N_0,J)$. The KK circle is now non-trivially 
fibered over the 4d spacetime with charge $N_0$. We now take a $N_0$-fold 
covering space of this solution whilst keeping the local geometry fixed in 
Planck units. This amounts to considering $N_0$ copies of our original 
black hole. The resulting solution has parameters $(0,1,N_0 J)$. The 
radius $R$ of the KK circle  increases to $N_0 R$ and the entropy is $N_0 
S$.

Next we apply the branch-exchange transformation to obtain a KK black hole with charges $(2N_0 J,1,0)$ and entropy $N_0 S$. Note that this is on the ergo-free branch.

Now we T-dualize on $T^6$ to obtain a KK black hole with charges $(1,2N_0 J,0)$ and then take a $K$-fold cover of the KK circle (where $K$ divides $2N_0 J$), keeping the local geometry fixed in Planck units. This gives a new black hole with parameters $(K^2,2N_0 J/K,0)$ and entropy $K N_0 S$.

In summary, we have explained why the entropy of our black hole should be 
$1/(KN_0)$ times that of an extremal KK black hole with parameters $(K^2,2 
N_0 J/K,0)$. For large $K,N_0J/K$, the entropy $2 \pi K N_0 J$ of the 
latter was reproduced by a statistical counting of states in 
\cite{EH}.\footnote{ This counting requires that $K^2 = 4k^3 N$ and $2N_0 
J/K = 4 l^3 N$ for integers $k,l,N$ with $N \gg 1$. We can arrange this 
e.g. by taking $K=2n$, $k=1$, $N=n^2$, $N_0 = l^3$, $J=4n^3$. Hence, for 
the ring, $J_1 = l^3+4n^3$, $J_2 = 4n^3$. This is a restriction on the 
charges of the original solution. Similar restrictions apply to 
\cite{EH,EM,HR}.} Hence this counting predicts an entropy $2\pi J$ for our 
original black ring. Setting $J=J_2$, this agrees with the 
Bekenstein-Hawking entropy (\ref{eqn:extremalring}).

\subsection{General ergo-branch black holes}

With slight modification, the above argument can also be applied to 
general extremal ergo-branch KK black holes in order to relax the 
condition (\ref{eqn:consistency}). Above we started with $N_6=0$ but now 
we consider an ergo-branch solution with parameters $(N_0,N_6,J)$ and 
$N_6>0$. Let $S$ denote the entropy.

First we go to a $N_6$-fold covering space of the KK circle keeping the 
local geometry fixed in Planck units. This gives us a KK black hole with 
parameters $(N_0 N_6^2,1,N_6J)$ and entropy $N_6 S$. Now perform a 
branch-exchange transformation. This gives an ergo-free extremal KK black 
hole with parameters $(2 N_6 J,1,N_0 N_6^2/2)$ and entropy $N_6 S$.

Next, T-dualize on $T^6$ to obtain a solution with parameters $(1,2N_6 
J,N_0 N_6^2/2)$, and taking a $K$-fold cover of the KK circle gives a 
solution with parameters $(K^2,2N_6 J/K,K N_0 N_6^2/2)$ and entropy $K N_6 
S$. This is a solution whose entropy was calculated microscopically in 
\cite{EH}\footnote{Again assuming $K^2 = 4 k^3 N$, $2N_6 J/K = 4 l^3 N$ 
for integers $k,l,N$ with $N \gg 1$.} with the result $2 \pi K N_6 
\sqrt{J^2 - N_0^2 N_6^2/4}$, so 
dividing by $K N_6$ exactly reproduces the Bekenstein-Hawking entropy 
(\ref{eqn:ergo}) of our original black hole.

\section{Discussion}

In this paper, we have presented a microscopic calculation of the entropy of extremal vacuum black rings. Our approach was based on the mapping from a black ring to a black string that has been succesful for BPS black rings \cite{benakraus,cyrier}. 
For BPS rings, this approach has several limitations, which were discussed in \cite{ringreview}. Similar limitations apply for vacuum rings. For example, since this approach cannot distinguish a black ring from a boosted black string, it provides no understanding of the lower bound (\ref{eqn:Jrange}) on $J_1$. Furthermore, since the calculation applies only to black rings, and not to asymptotically flat Myers-Perry black holes, it provides no hint of what distinguishes a black ring from a topologically spherical black hole at the microscopic level. 

\medskip

\centerline{\bf Acknowledgments}

\medskip

\noindent I am grateful to Henriette Elvang, Roberto Emparan and Hari 
Kunduri for comments on a 
draft of this paper. This work was supported by the Royal Society.


\begin{thebibliography}{99}

\bibitem{EH}
R.~Emparan and G.~T.~Horowitz,
  ``Microstates of a neutral black hole in M theory,''
  Phys.\ Rev.\ Lett.\  {\bf 97}, 141601 (2006)
  [arXiv:hep-th/0607023].

\bibitem{EM}
R.~Emparan and A.~Maccarrone,
  ``Statistical description of rotating Kaluza-Klein black holes,''
  Phys.\ Rev.\  D {\bf 75}, 084006 (2007)
  [arXiv:hep-th/0701150].

\bibitem{HR}
G.~T.~Horowitz and M.~M.~Roberts,
  ``Counting the Microstates of a Kerr Black Hole,''
  arXiv:0708.1346 [hep-th].

\bibitem{senreview}
A.~Sen,
  ``Black Hole Entropy Function, Attractors and Precision Counting of
  Microstates,''
  arXiv:0708.1270 [hep-th].

\bibitem{rotatt}
D.~Astefanesei, K.~Goldstein, R.~P.~Jena, A.~Sen and S.~P.~Trivedi,
  ``Rotating attractors,''
  JHEP {\bf 0610}, 058 (2006)
  [arXiv:hep-th/0606244].

\bibitem{agmdab}
  D.~Astefanesei, K.~Goldstein and S.~Mahapatra,
  ``Moduli and (un)attractor black hole thermodynamics,''
  arXiv:hep-th/0611140;
A.~Dabholkar, A.~Sen and S.~P.~Trivedi,
  ``Black hole microstates and attractor without supersymmetry,''
  JHEP {\bf 0701}, 096 (2007)
  [arXiv:hep-th/0611143].

\bibitem{ER}
R.~Emparan and H.~S.~Reall,
  ``A rotating black ring in five dimensions,''
  Phys.\ Rev.\ Lett.\  {\bf 88}, 101101 (2002)
  [arXiv:hep-th/0110260].

\bibitem{PS}
A.~A.~Pomeransky and R.~A.~Sen'kov,
  ``Black ring with two angular momenta,''
  arXiv:hep-th/0612005.

\bibitem{henriette}
H.~Elvang and M.~J.~Rodriguez,
  ``Bicycling Black Rings,''
  arXiv:0712.2425 [hep-th].

\bibitem{KLR}
H.~K.~Kunduri, J.~Lucietti and H.~S.~Reall,
 ``Near-horizon symmetries of extremal black holes,''
 Class.\ Quant.\ Grav.\  {\bf 24}, 4169 (2007)
 [arXiv:0705.4214 [hep-th]].

\bibitem{bpsrings}
H.~Elvang, R.~Emparan, D.~Mateos and H.~S.~Reall,
  ``A supersymmetric black ring,''
  Phys.\ Rev.\ Lett.\  {\bf 93}, 211302 (2004)
  [arXiv:hep-th/0407065],
``Supersymmetric black rings and three-charge supertubes,''
  Phys.\ Rev.\  D {\bf 71}, 024033 (2005)
  [arXiv:hep-th/0408120];
  I.~Bena and N.~P.~Warner,
  ``One ring to rule them all ... and in the darkness bind them?,''
  Adv.\ Theor.\ Math.\ Phys.\  {\bf 9}, 667 (2005)
  [arXiv:hep-th/0408106];
J.~P.~Gauntlett and J.~B.~Gutowski,
  ``General concentric black rings,''
  Phys.\ Rev.\  D {\bf 71}, 045002 (2005)
  [arXiv:hep-th/0408122].

\bibitem{benakraus}
I.~Bena and P.~Kraus,
  ``Microscopic description of black rings in AdS/CFT,''
  JHEP {\bf 0412}, 070 (2004)
  [arXiv:hep-th/0408186].

\bibitem{cyrier}
M.~Cyrier, M.~Guica, D.~Mateos and A.~Strominger,
  ``Microscopic entropy of the black ring,''
  Phys.\ Rev.\ Lett.\  {\bf 94}, 191601 (2005)
  [arXiv:hep-th/0411187].

\bibitem{dipole}
R.~Emparan,
  ``Rotating circular strings, and infinite non-uniqueness of black rings,''
  JHEP {\bf 0403}, 064 (2004)
  [arXiv:hep-th/0402149].

\bibitem{emparannew}
R.~Emparan, ``Exact
microscopic entropy of non-supersymmetric extremal black rings", to 
appear.

\bibitem{myersperry}
R.~C.~Myers and M.~J.~Perry,   
  ``Black Holes In Higher Dimensional Space-Times,''
  Annals Phys.\  {\bf 172}, 304 (1986).

\bibitem{KKBH}
D.~Rasheed,
  ``The Rotating dyonic black holes of Kaluza-Klein theory,''
  Nucl.\ Phys.\  B {\bf 454}, 379 (1995)
  [arXiv:hep-th/9505038].
F.~Larsen,
  "Rotating Kaluza-Klein black holes,''
  Nucl.\ Phys.\  B {\bf 575}, 211 (2000)
  [arXiv:hep-th/9909102].

\bibitem{ringreview}
R.~Emparan and H.~S.~Reall,
  ``Black rings,''
  Class.\ Quant.\ Grav.\  {\bf 23}, R169 (2006)
  [arXiv:hep-th/0608012].

\end{thebibliography}
\end{document}